\documentclass[12pt]{article}

\newcommand{\bc}{\begin{center}}
\newcommand{\ec}{\end{center}}
\newcommand{\bd}{\begin{displaymath}}
\newcommand{\ed}{\end{displaymath}}
\newcommand{\be}{\begin{equation}}
\newcommand{\ee}{\end{equation}}
\newcommand{\ba}{\begin{array}}
\newcommand{\ea}{\end{array}}
\newcommand{\bt}{\begin{tabular}}
\newcommand{\et}{\end{tabular}}

\begin{document}
\title{\bf Dark matter from encapsulated atoms}

\author{C.D.~Froggatt${}^{1,2}$ and H.B.~Nielsen${}^{2}$
\\[15mm] \itshape{${}^{1}$ Department of
Physics and Astronomy,}\\[3mm] \itshape{Glasgow University,
Glasgow, Scotland}\\[3mm] \itshape{${}^{2}$ The Niels Bohr
Institute, Copenhagen, Denmark}\\[3mm]
}

\date{}

\maketitle

\thispagestyle{empty}

\setcounter{page}{0}

\begin{abstract}
We propose that dark matter consists of collections of atoms
encapsulated inside pieces of an alternative vacuum, in which the
Higgs field vacuum expectation value is appreciably smaller than
in the usual vacuum. The alternative vacuum is supposed to have
the same energy density as our own. Apart from this degeneracy of
vacuum phases, we do not introduce any new physics beyond the
Standard Model. The dark matter balls are estimated to have a
radius of order 20 cm and a mass of order $10^{11}$ kg. However
they are very difficult to observe directly, but inside dense
stars may expand eating up the star and cause huge explosions
(gamma ray bursts). The ratio of dark matter to ordinary
baryonic matter is estimated to be of the order of the ratio
of the binding energy per nucleon in helium to the difference
between the binding energies per nucleon in heavy nuclei and in
helium. Thus we predict approximately five times as much dark
matter as ordinary baryonic matter!
\end{abstract}

\newpage

\section{Introduction}

Recent ``precision" cosmological measurements agree on a so-called
concordant model (see the Reviews of Astrophysics and Cosmology in
\cite{pdg}), according to which the Universe is flat with
$\Omega$, the ratio of its energy density to the critical density,
being very close to unity. The energy budget of the Universe is
presently dominated by three components: ordinary baryonic matter
($\Omega_{ordinary} \simeq 0.04$), dark matter ($\Omega_{dark}
\simeq 0.23$) and dark energy ($\Omega_{\Lambda} \simeq 0.73$).
The main evidence for the density of ordinary matter comes from
the abundances of the light elements formed in the first three
minutes by big bang nucleosynthesis (BBN). The evidence for the
dark matter density comes from galactic rotation curves, motions
of galaxies within clusters, gravitational lensing and analyses
(e.g.~WMAP \cite{spergel}) of the cosmic microwave background
radiation. The need for a form of dark energy, such as a tiny
cosmological constant $\Lambda$, is provided by the evidence for
an accelerating Universe from observations of type Ia supernovae,
large scale structure and the WMAP data.

In this paper we shall concentrate on the dark matter component.
It must be very stable, with a lifetime greater than $10^{10}$
years. The dark matter density is of a similar order of magnitude
as that of ordinary matter, with a ratio of \be
\frac{\Omega_{dark}}{\Omega_{ordinary}} \simeq 6 \ee Also the dark
matter was non-relativistic at the time of the onset of galaxy
formation (i.e.~cold dark matter).

According to folklore, no known elementary particle can account
for all of the dark matter. Many hypothetical particles have been
suggested as candidates for dark matter, of which the most popular
is the lightest supersymmetric particle (LSP): the neutralino. The
stability of the the LSP is imposed by the assumption of R-parity
conservation. The LSP density is predicted to be close to the
critical density for a heavy neutralino \cite{jkg} with mass
$m_{LSP} \sim 100-1000$ GeV, but {\it a priori} it is unrelated to
the density of normal matter.

However we should like to emphasize that the dark matter could in
fact be baryonic, if it were effectively separated from normal
matter at the epoch of BBN. This separation must therefore already
have been operative 1 second after the big bang, when the
temperature was of order 1 MeV. Our basic idea is that dark matter
consists of ``small balls" of an alternative Standard Model vacuum
degenerate with the usual one, containing close-packed nuclei and
electrons and surrounded by domain walls separating the two vacua
\cite{prl}. The baryons are supposed to be kept inside the balls
due to the vacuum expectation value (VEV) of the Weinberg-Salam
Higgs field $<\phi_{WS}>$ being smaller, say by a factor of 2, in
the alternative phase. The quark and lepton masses \be m_f = g_f
<\phi_{WS}> \ee are then reduced (by a factor of 2). We use an
additive quark mass dependence approximation for the nucleon mass
\cite{weinberg}: \be m_N = m_0 + \sum_{i=1}^3 m_{q_i}, \ee where
the dominant contribution $m_0$ to the nucleon mass arises from
the confinement of the quarks. Then, assuming quark masses in our
phase of order $m_u \sim 5$ MeV and $m_d \sim 8$ MeV, we obtain a
reduction in the nucleon mass in the alternative phase by an
amount $\Delta m_N \sim 10$ MeV. The pion may be considered as a
pseudo-Goldstone boson with a mass squared proportional to the sum
of the masses of its constituent quarks: \be M_{\pi}^2 \propto m_u
+ m_d. \ee It follows that the pion mass is also reduced (by a
factor of $\sqrt{2}$) in the alternative phase. The range of the
pion exchange force is thereby increased and so the nuclear
binding energies are larger in the alternative phase, by an amount
comparable to the binding they already have in normal matter. We
conclude it would be energetically favourable for the dark matter
baryons to remain inside balls of the alternative vacuum for
temperatures lower than about 10 MeV. These dark matter nucleons
would be encapsulated by the domain walls, remaining relatively
inert and not disturbing the successful BBN calculations in our
vacuum. We should note that a model for dark matter using an
alternative phase in QCD has been proposed by Oaknin and
Zhitnitsky \cite{OZ}.

\section{Degenerate vacua in the Standard Model}

The existence of another vacuum could be due to some genuinely new
physics, but here we want to consider a scenario, which does not
introduce any new fundamental particles or interactions beyond the
Standard Model. Our main assumption is that the dark energy or
cosmological constant $\Lambda$ is not only fine-tuned to be tiny
for one vacuum but for several, which we have called
\cite{bnp,origin,MPP,bn,fn2} the Multiple Point Principle (MPP).
This entails a fine-tuning of the parameters (coupling constants)
of the Standard Model analogous to the fine-tuning of the
intensive variables temperature and pressure at the triple point
of water, due to the co-existence of the three degenerate phases:
ice water and vapour.

Different vacuum phases can be obtained by having different
amounts of some Bose-Einstein condensate. We are therefore led to
consider a condensate of a bound state of some SM particles.
Indeed, in this connection, we have previously proposed
\cite{itep,portoroz,coral,pascos04} the existence of a new exotic
strongly bound state made out of 6 top quarks and 6 anti-top
quarks. The reason that such a bound state had not been considered
previously is that its binding is based on the collective effect
of attraction between several quarks due to Higgs exchange. In
fact our calculations show that the binding could be so strong
that the bound state is on the verge of becoming tachyonic and
could form a condensate in an alternative vacuum degenerate with
our own. With the added assumption of a third Standard Model
phase, having a Higgs vacuum expectation value of the order of the
Planck scale, we obtained a value of 173 GeV for the top quark
mass \cite{fn2} and even a solution of the hierarchy problem, in
the sense of obtaining a post-diction of the order of magnitude of
the ratio of the weak to the Planck scale
\cite{itep,portoroz,coral,pascos04}. However this third Planck
scale vacuum is irrelevant for our dark matter scenario.

With the existence of just the 2 degenerate vacua domain walls
would have easily formed, separating the different phases of the
vacuum occurring in different regions of space, at high enough
temperature in the early Universe. Since we assume the weak scale
physics of the top quark and Higgs fields is responsible for
producing these bound state condensate walls, their energy scale
will be of order the top quark mass. We note that, unlike walls
resulting from the spontaneous breaking of a discrete symmetry,
there is an asymmetry between the two sides of the the wall. So,
in principle, a wall can readily contract to one side or the other
and disappear.

\section{Formation of dark matter balls in the early Universe}

We now describe our favoured scenario for how the dark matter
balls formed. Let us denote the order parameter field describing
the new bound state which condenses in the alternative phase by
$\phi_{NBS}$. In the early Universe it would fluctuate
statistically mechanically and, as the temperature $T$ fell below
the weak scale, would have become more and more concentrated
around the -- assumed equally deep -- minima of the effective
potential $V_{eff}(\phi_{NBS})$. There was then an effective
symmetry between the vacua, since the vacua had approximately the
same free energy densities. So the two phases would have formed
with comparable volumes separated by domain walls. Eventually the
small asymmetry between their free energy densities would have led
to the dominance of one specific phase inside each horizon region
and, finally, the walls would have contracted away. However it is
a very detailed dynamical question as to how far below the weak
scale the walls would survive. It seems quite possible that they
persisted until the temperature of the Universe fell to around 1
MeV.

We imagine that the disappearance of the walls in our phase --
except for very small balls of the fossil phase -- occurred when
the temperature $T$ was of the order of 1 MeV to 10 MeV. During
this epoch the collection of nucleons in the alternative phase was
favoured by the Boltzmann factor $\exp(-\Delta m_N/T)$. Thus the
nucleons collected more and more strongly into the alternative
phase, leaving relatively few nucleons outside in our phase. We
suppose that a rapid contraction of the alternative phase set in
around a temperature $T \sim 1$ MeV.

Due to the higher density and stronger nuclear binding,
nucleosynthesis occurred first in the alternative phase. Ignoring
Coulomb repulsion, the temperature $T_{NUC}$ at which a given
species of nucleus with nucleon number A is thermodynamically
favoured is given \cite{kolbturner} by:
\begin{equation}
T_{NUC} = \frac{B_A /(A-1)}{\ln(\eta^{-1}) + 1.5 \ln(m_N
/T_{NUC})}.
\end{equation}
Here $B_A$ is the binding energy of the nucleus -- in the phase in
question of course -- $\eta$= $\frac{n_B}{n_{\gamma}}$ is the
ratio of the baryon number density relative to the photon density,
and $m_N$ is the nucleon mass. In our phase, for example, the
temperature for ${^4}$He to be thermodynamically favoured turns
out from this formula to be 0.28 MeV. In the other phase, where
the Higgs field has a lower VEV by a factor of order unity, the
binding energy $ B_A$ is bigger and, with say $\eta \sim 10^{-3}$,
$^4$He could have been produced at $T \sim 1$ MeV.

We assume that the alternative phase continued to collect up any
nucleons from our phase and that, shortly after $^4$He production,
there were essentially no nucleons left in our phase. The rapid
contraction of the balls continued until there were more nucleons
than photons, $\eta > 1$, in the alternative phase and fusion to
heavier nuclei, such as $^{12}$C and $^{56}$Fe, took place, still
with $T \sim 1$ MeV. A chain reaction could then have been
triggered, resulting in the explosive heating of the whole ball as
the $^4$He burnt into heavier nuclei. The excess energy would have
been carried away by nucleons freed from the ball.

At this stage of internal fusion, the balls of the alternative
phase would have been so small that any nucleons in our phase
would no longer be collected into the balls. So the nucleons
released by the internal fusion would stay forever outside the
balls and make up {\it normal matter}. This normal matter then
underwent the usual BBN in our phase.

\section{Prediction of the ratio of dark matter to normal matter}

According to the above internal fusion scenario, the ratio of the
normal matter density to the total matter density is given by: \be
\frac{\Omega_{ordinary}}{\Omega_{matter}} = \frac{\mbox{Number of
nucleons released}}{\mbox{Total number of nucleons}} \ee The
fraction of nucleons released from the balls of alternative phase
during the internal fusion can be obtained from a simple energy
conservation argument.

Before the further internal fusion process took place, the main
content of the balls was in the form of ${^4}$He nuclei. Now the
nucleons in a ${^4}$He nucleus have a binding energy of 7.1 MeV in
normal matter in our phase, while a typical ``heavy'' nucleus has
a binding energy of 8.5 MeV for each nucleon \cite{ring}. Let us,
for simplicity, assume that the ratio of these two binding
energies per nucleon is the same in the alternative phase and use
the normal binding energies in our estimate below. Thus we take
the energy released by the fusion of the helium into heavier
nuclei to be 8.5 MeV - 7.1 MeV = 1.4 MeV per nucleon. Now we can
calculate what fraction of the nucleons, counted as {\it a priori}
initially sitting in the heavy nuclei, can be released by this 1.4
MeV per nucleon. Since they were bound inside the nuclei by 8.5
MeV relative to the energy they would have outside, the fraction
released should be (1.4 MeV)/(8.5 MeV) = $0.16_5$ = 1/6. So we
predict that the normal baryonic matter should make up 1/6 of the
total amount of matter, dark as well as normal baryonic. According
to astrophysical fits \cite{spergel}, giving 23\% dark matter and
4\% normal baryonic matter relative to the critical density, the
amount of normal baryonic matter relative to the total matter is
$\frac{4\%}{23\% + 4\%} = 4/27 = 0.15$. This is in remarkable
agreement with our prediction.

\section{Properties of dark matter balls}

The size of the balls depends sensitively on the order of
magnitude assumed for the wall energy density, which we take to be
of the weak scale or about 100 GeV. Let us first consider the
stability condition for these balls. For a ball of radius R, the
wall tension $s$ is given by \be s \approx (100\ \mathrm{GeV})^3
\ee which provides a pressure $\frac{s}{R}$ that must be balanced
by the electron pressure. The energy needed to release a nucleon
from the alternative vacuum into our vacuum is approximately 10
MeV. So the maximum value for the electron Fermi level inside the
balls is $\sim 10$ MeV, since otherwise it would pay for electrons
and associated protons to leave the alternative vacuum. Thus the
maximum electron pressure is of order (10 MeV)$^4$.

In order that the pressure from the wall should not quench this
maximal electron pressure, we need to satisfy the stability
condition: \be s/R = \frac{(100\ \mathrm{GeV})^3}{R} < (10\
\mathrm{MeV})^4 = 10^{-8}\ \mathrm{GeV}^4. \ee This means the ball
radius must be larger than a critical radius given by: \be R
> R_{crit} = 10^{14}\ \mathrm{GeV}^{-1} = 2\ \mathrm{cm}. \ee
If the balls have a radius smaller than $R_{crit}$, they will
implode. These critical size balls have a nucleon number density
of \be n_e = (10\ \mathrm{MeV})^3 = \frac{1}{(20\ \mathrm{fm})^3}
\simeq 10^{35}\ \mathrm{cm}^{-3}. \ee  So, with $R_{crit} = 2$ cm,
it contains of order $N_e \simeq 10^{36}$ electrons and
correspondingly of order $N_B \simeq 10^{36}$ baryons, with a mass
of order $M_B \simeq 10^9$ kg.

We estimate the typical radius of a dark matter ball in our
scenario to be of order 20 cm. It contains of order $N_B = 3\times
10^{37}$ baryons and has a mass of order $M_B = 10^{11}$ kg =
$10^{-19}M_{\odot} = 10^{-14}M_{\oplus}$. Therefore dark matter
balls can not be revealed by microlensing searches, which are only
sensitive to massive astrophysical compact objects with masses
greater than $10^{-7}M_{\odot}$ \cite{afonso}. Since the dark
matter density is 23\% of the critical density
$\rho_{\mathrm{crit}} = 10^{-26}$ kg/m$^3$, a volume of about
$10^{37}$ m$^3$ = (20 astronomical units)$^3$ will contain on the
average just one dark matter ball.

Assuming the sun moves with a velocity of 100 km/s relative to the
dark matter and an enhanced density of dark matter in the galaxy
of order $10^5$ higher than the average, the sun would hit of
order $10^8$ dark matter balls of total mass $10^{19}$ kg in the
lifetime of the Universe. A dark matter ball passing through the
sun would plough through a mass of sun material similar to its own
mass. It could therefore easily become bound into an orbit say or
possibly captured inside the sun, but be undetectable from the
earth. On the other hand, heavy stars may capture some dark matter
balls impinging on them.

In the lifetime of the Universe, the earth would hit $10^4$ or so
dark matter balls. However they would have gone through the earth
without getting stopped appreciably. It follows that DAMA
\cite{dama} would not have any chance of seeing our dark matter
balls, despite their claim to have detected a signal for dark
matter in the galactic halo. However EDELWEISS \cite{edelweiss},
CRESST \cite{cresst} and CDMS \cite{cdms} do not confirm the
effect seen by DAMA. It is also possible that DAMA saw something
other than dark matter. Geophysical evidence for the dark matter
balls having passed through the earth would also be extremely
difficult to find.

We conclude that the dark matter balls are very hard to see
directly. On the other hand, we could imagine that dark matter
balls had collected into the interior of a collapsing star. Then,
when the density in the interior of the star gets sufficiently
big, the balls could be so much disturbed that they would explode.
The walls may then start expanding into the dense material in the
star, converting part of the star to dark matter. As the wall
expands the pressure from the surface tension diminishes and lower
and lower stellar density will be sufficient for the wall to be
driven further out through the star material. This could lead to
releasing energy of the order of 10 MeV per nucleon in the star,
which corresponds to of the order of one percent of the Einstein
energy of the star! Such events would give rise to really huge
energy releases, perhaps causing supernovae to explode and
producing the canonballs suggested by Dar and De Rujula
\cite{deRujula} to be responsible for the cosmic gamma ray bursts.
We should note that a different (SUSY) phase transition inside the
star has already been suggested \cite{clavelli} as an explanation
for gamma ray bursts.

A dark matter ball can also explode due to the implosion of its
wall. Such an implosive instability might provide a mechanism for
producing ultra high energy cosmic rays from seemingly empty
places in the Universe. This could help to resolve the
Greisen-Zatsepin-Kuzmin \cite{G,ZK} cut-off problem.

\section{Conclusion}

Under the assumption that there be at least two different phases
of the vacuum with very closely the same tiny energy density or
cosmological constant, we have put forward an idea for what dark
matter could be. Indeed we suggest that dark matter consists of
baryons hidden inside pieces of an alternative vacuum with a
smaller Higgs field VEV. The SM might provide such a second vacuum
degenerate with our own, due to the condensation of an exotic $6t
+ 6\overline{t}$ strongly bound state. The ratio of dark matter to
ordinary matter is expressed as a ratio of nuclear binding
energies and predicted to be about 5. Big bang nucleosynthesis is
supposed to proceed as usual in our vacuum relatively undisturbed
by the crypto-baryonic dark matter encapsulated in a few balls of
the alternative vacuum.

We estimate that a typical dark matter ball has a radius of about
20 cm and a mass of order $10^{11}$ kg. The dark matter balls are
very difficult to detect directly, but they might be responsible
for gamma ray bursts or ultra high energy cosmic rays.

\section*{Acknowledgements}

We acknowledge discussions with D.~Bennett and Y.~Nagatani in the
early stages of this work. CDF would like to acknowledge the hospitality
of the Niels Bohr Institute and support from the Niels Bohr Institute
Fund and PPARC.

\end{document}